\documentclass[reprint,amsmath,amssymb,aps,prl,floatfix]{revtex4-1}  

\usepackage{graphicx}
\usepackage{dcolumn}
\usepackage{bm}
\usepackage{hyperref}
\hypersetup{colorlinks=True,citecolor=blue,linkcolor=blue,urlcolor=blue}
\usepackage{multirow} 
\usepackage{enumerate}

\def\be{\begin{equation}} 
\def\ee{\end{equation}} 
\def\intd{\,\mathrm{d}} 

\begin{document}

\title{Spontaneous Transport Barriers Quench Turbulent Resistivity in 2D MHD}
\author{Xiang Fan}\affiliation{University of California at San Diego, La Jolla, California 92093}
\author{P. H. Diamond}\affiliation{University of California at San Diego, La Jolla, California 92093}
\author{L. Chac\'on}\affiliation{Los Alamos National Laboratory, Los Alamos, New Mexico 87545}
\date{\today} 

\begin{abstract}
This Letter identifies the physical mechanism for the quench of turbulent resistivity in 2D MHD. Without an imposed, ordered magnetic field, a multi-scale, blob-and-barrier structure of magnetic potential forms spontaneously. Magnetic energy is concentrated in thin, linear barriers, located at the interstices between blobs. The barriers quench the transport and kinematic decay of magnetic energy. The local transport bifurcation underlying barrier formation is linked to the inverse cascade of $\langle A^2\rangle$ and negative resistivity, which induce local bistability. For small scale forcing, spontaneous layering of the magnetic potential occurs, with barriers located at the interstices between layers. This structure is effectively a magnetic staircase.
\end{abstract}

\maketitle

\section{Introduction}

The evolution of mean quantities in turbulence is frequently modelled as a \textit{transport process}, using ideas from the kinetic theory of gases. A classic example is that of Prandtl's theory of turbulent boundary layers, which first proposed the use of an eddy viscosity - based upon mixing length theory - to calculate mean flow profiles at high Reynolds number. Magnetohydrodynamics (MHD) presents additional challenges, especially at high magnetic Reynolds number $\mathrm{Rm}$. There, models based on transport theory concepts are central to our understanding of mean B ($\langle\mathbf{B}\rangle$) evolution in turbulent flows. Indeed, the well-known theory of mean field electrodynamics (Moffatt \cite{moffatt_magnetic_1983}) employs transport coefficients $\alpha$, $\beta$ - related to turbulent helicity and energy, respectively - to describe the growth and transport of a mean magnetic field. Such models are heavily utilized in dynamo theory - the study of how large scale fields are formed. The turbulent or ``eddy'' \textit{resistivity}, $\eta_T$, is ubiquitous in these models (and corresponds to $\beta$ above). While $\eta_T$ is often taken as kinematic ($\eta_T\sim\eta_K\sim\sum_{\mathbf{k}}\langle\tilde{v}^2\rangle_{\mathbf{k}}\tau_c$ where $\tau_c$ is the self-correlation time) for many applications, nonlinear dependence of $\eta_T$ on magnetic field and potential has been observed in numerous simulations \cite{cattaneo_suppression_1991,vainshtein_turbulent_1991,vainshtein_nonlinear_1992,cattaneo_effects_1994,balmforth_dynamics_1998,lazarian_reconnection_1999,biskamp_two-dimensional_2001,field_dynamical_2002,mininni_numerical_2005,kim_consistent_2006,silvers_dynamic_2005,silvers_choice_2006,kleeorin_nonlinear_2007,keating_turbulent_2007,keating_cross-phase_2008,keating_turbulent_2008,tobias_-plane_2007,eyink_fast_2011,kondic_decay_2016,mak_vortex_2017}. Such nonlinearity arises from the fact that the magnetic fields alter the turbulent flows which scatter them. As this nonlinearity tends to reduce $\eta_T$ relative to kinematic expectations, such trends are referred to as quenching. $\mathrm{Rm}$ dependent quenching -- i.e. when the product $\mathrm{Rm}\langle\mathbf{B}\rangle^2$ enters -- is of particular interest, as it signals that for the relevant case of high $\mathrm{Rm}$, relatively weak fields can produce significant feedback on transport and evolution processes. Such $\mathrm{Rm}$-dependent feedback has been associated with Alfvenization (i.e. the conversion of hydrodynamics turbulence to Alfven wave turbulence) and/or with the balance of magnetic helicity $\langle\mathbf{A}\cdot\mathbf{B}\rangle$ (i.e. in 3D) or $\langle A^2\rangle$ (i.e. in 2D). Both arguments ultimately point to \textit{memory}, due to the freezing-in law, as the origin of the quench. The quenching problem is also relevant to models of fast reconnection and impulsive energy release processes in MHD, as it constrains the size of (frequently invoked) anomalous dissipation \cite{xi_phase_2014,xi_impact_2014}. More generally, it is an important paradigm of the transport of an active scalar.

\begin{figure}[htbp] 
    \centering
    \includegraphics[width=0.9\columnwidth]{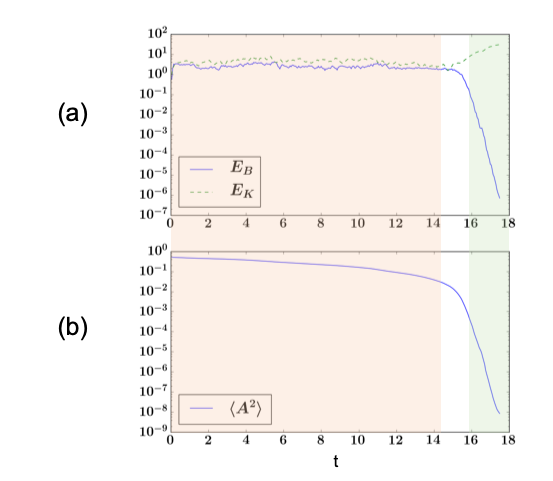}
    \caption{Time evolution of (a) magnetic energy $E_B$ and kinetic energy $E_K$; (b) $\langle A^2\rangle$ in Run1. The suppression stage is marked in orange, and the kinematic decay stage in green. The decay of $E_B$ is slow in the suppression stage, which is consistent with previous studies. The decay of $\langle A^2\rangle$ is also slow in the suppression stage, and is more smooth compared to $E_B$.}
    \label{energy_time}
\end{figure}

In a seminal paper \cite{cattaneo_suppression_1991} which broached the quenching question, Cattaneo and Vainshtein (CV) presented numerical simulations of 2D MHD turbulence which demonstrated that even a weak large scale magnetic field is sufficient to quench the turbulent transport of the active scalar $A$ (the magnetic potential). Based on ideas from mean field theory, CV suggested -- and presented simulations to support -- the idea that $\eta_T$ is given by
\be 
\eta_T\sim \frac{\langle v^2\rangle^{1/2} l}{1+\frac{1}{\mu_0\rho}\mathrm{Rm}\langle \mathbf{B}\rangle^2/\langle v^2\rangle}\label{eta_T_eqn}
\ee 
The mean field $\langle \mathbf{B}\rangle$ here is estimated using:
\be 
|\langle\mathbf{B}\rangle|\sim\sqrt{\langle A^2\rangle}/L_0\label{b_avg_eqn}
\ee 
where $L_0$ is system size. For $\frac{1}{\mu_0\rho}\mathrm{Rm}\langle \mathbf{B}\rangle^2/\langle v^2\rangle<1$, $\eta_T\sim\eta_K\sim\langle v^2\rangle^{1/2} l$.  While for $\frac{1}{\mu_0\rho}\mathrm{Rm}\langle \mathbf{B}\rangle^2/\langle v^2\rangle>1$, $\eta_T\ll\eta_K$, so $\eta_T$ is quenched. It is important to note that, in view of Cowling's Theorem, suppression occurs only for a time of \textit{limited duration}, without external forcing of $A$. After the afore mentioned suppression stage, rapid decay of the magnetic field occurs, and $\eta_T$ reverts to $\eta_K$. The evolutions of $E_B$, $E_K$ (magnetic and kinetic energy) and $\langle A^2\rangle$ (mean square potential) are shown in Fig.~\ref{energy_time} (a, b).

Equation (\ref{eta_T_eqn}) was also obtained analytically from statistical theory, assuming the presence of an imposed weak large scale field $\mathbf{B}_0$ (i.e. $\langle\mathbf{B}\rangle=\mathbf{B}_0$) \cite{gruzinov_self-consistent_1994,gruzinov_nonlinear_1996,diamond_self-consistent_2005,diamond_modern_2010}. (Note the assumptions that $|\langle\mathbf{B}\rangle|$ is determined by root-mean-square $A$ and the system size in CV.) The derivation made use of $\langle A^2\rangle$ balance to constrain the turbulent resistivity \cite{zeldovich_magnetic_1957,pouquet_strong_1976,pouquet_two-dimensional_1978}. $\mathrm{Rm}$-dependence of the quench stems from the fact that $\langle A^2\rangle$ is conserved up to resistive diffusion. This early work on resistivity quenching triggered a tidal wave of subsequent studies of nonlinear dynamo evolution and quenching.

In this Letter, we show that, without an imposed, ordered magnetic field, $\mathrm{Rm}$-dependent quenching is intrinsically an intermittency phenomena, and can occur where a global mean field $\langle\mathbf{B}\rangle$ simply does not exist. Rather, turbulent resistivity quenching occurs due to intermittent \textit{transport barriers}. A transport barrier is a localized region of mixing and transport significantly lower than the mean thereof, i.e. $\eta_{T, local}<\bar{\eta}_T$. These barriers are extended, thin, linear features, into which strong $\langle B^2\rangle$ is concentrated. The barriers are formed by the $\langle B^2\rangle$ feedback on scalar transport, specifically by magnetic flux coalescence. Thus, transport quenching is manifestly not a mean field effect, as the structure of the field is more akin to a random network than to a smooth mean field. The barriers form in the interstices between blobs of $\langle A^2\rangle$, which are formed by the inverse cascade of $\langle A^2\rangle$. Overall, the magnetic potential and field have a structure of ``blob-and-barrier" at large $\mathrm{Rm}$, as shown in Fig.~\ref{snapshots}. In contrast to the assumptions of CV, the magnetic field exhibits \textit{two} non-trivial scales, i.e. the blob size $L_{blob}$ and the barrier width $W$, where $W\ll L_{blob}$. $L_{blob}$ characterizes the magnetic potential while $W$ characterizes the field intensity.

The $A$ field in the blob-and-barrier structure of 2D MHD resembles the concentration contrast field $\psi$ in the Cahn-Hilliard Navier-Stokes (CHNS) system \cite{ruiz_turbulence_1981,fan_cascades_2016,fan_formation_2017,fan_chns:_2018,pandit_overview_2017}. 

\section{Analysis: global}

In this Letter, the 2D MHD equations are solved using direct numerical simulation \cite{chacon_implicit_2002,chacon_2d_2003} with doubly periodic boundary condition: 
\begin{align}
\partial_t A+\mathbf{v}\cdot\nabla A&=\eta\nabla^2 A \label{MHD1}\\
\partial_t\omega+\mathbf{v}\cdot\nabla\omega&=\frac{1}{\mu_0\rho}\mathbf{B}\cdot\nabla\nabla^2 A+\nu\nabla^2\omega+f \label{MHD2}
\end{align}

Here $\omega$ is vorticity, $\eta$ is resistivity, $\nu$ is viscosity, $\mu_0\rho$ is magnetic permeability and density, and $f$ is an isotropic homogeneous external forcing, with wave number $k$ and magnitude $f_0$. The simulation box size is $L_0^2=1.0\times 1.0$ with $1024 \times 1024$ resolution. The parameters used are summarized in Table.~\ref{parameter_table}. The initial condition for the $\omega$ field is $\omega_I=0$ everywhere; the initial condition for $A$ field is a cosine function in Run1: $A_I(x,y) = A_0\cos{2\pi x}$. The setup of Run1 differs from that of Ref.~\cite{cattaneo_suppression_1991} only in the range of $\mathrm{Rm}$ studied.

\begin{table}[t]
\caption{Initial conditions, $k$ and $\mathrm{Rm}$ for the suppression stage. For all runs, $A_0=1.0$ and $f_0=30$.}
\begin{center}
\begin{tabular}{ccccccc}
\hline\hline
Runs & Initial Condition  & $\eta$ & $\nu$ & $1/(\mu_0\rho)$ & $k$ & Rm \\
\hline
Run1 & Bimodal & $1*10^{-4}$ & $1*10^{-4}$ & $0.04$ & $5$ & $\sim 500$ \\
Run2 & Unimodal & $1*10^{-4}$ & $1*10^{-4}$ & $0.04$ & $5$ & $\sim 500$ \\
Run3 & Bimodal & $1*10^{-4}$ & $2*10^{-3}$ & $0.01$ & $32$ & $\sim 150$ \\
\hline\hline
\end{tabular}
\end{center}
\label{parameter_table}
\end{table}%

\begin{figure*}[htbp] 
    \centering
    \includegraphics[width=\textwidth]{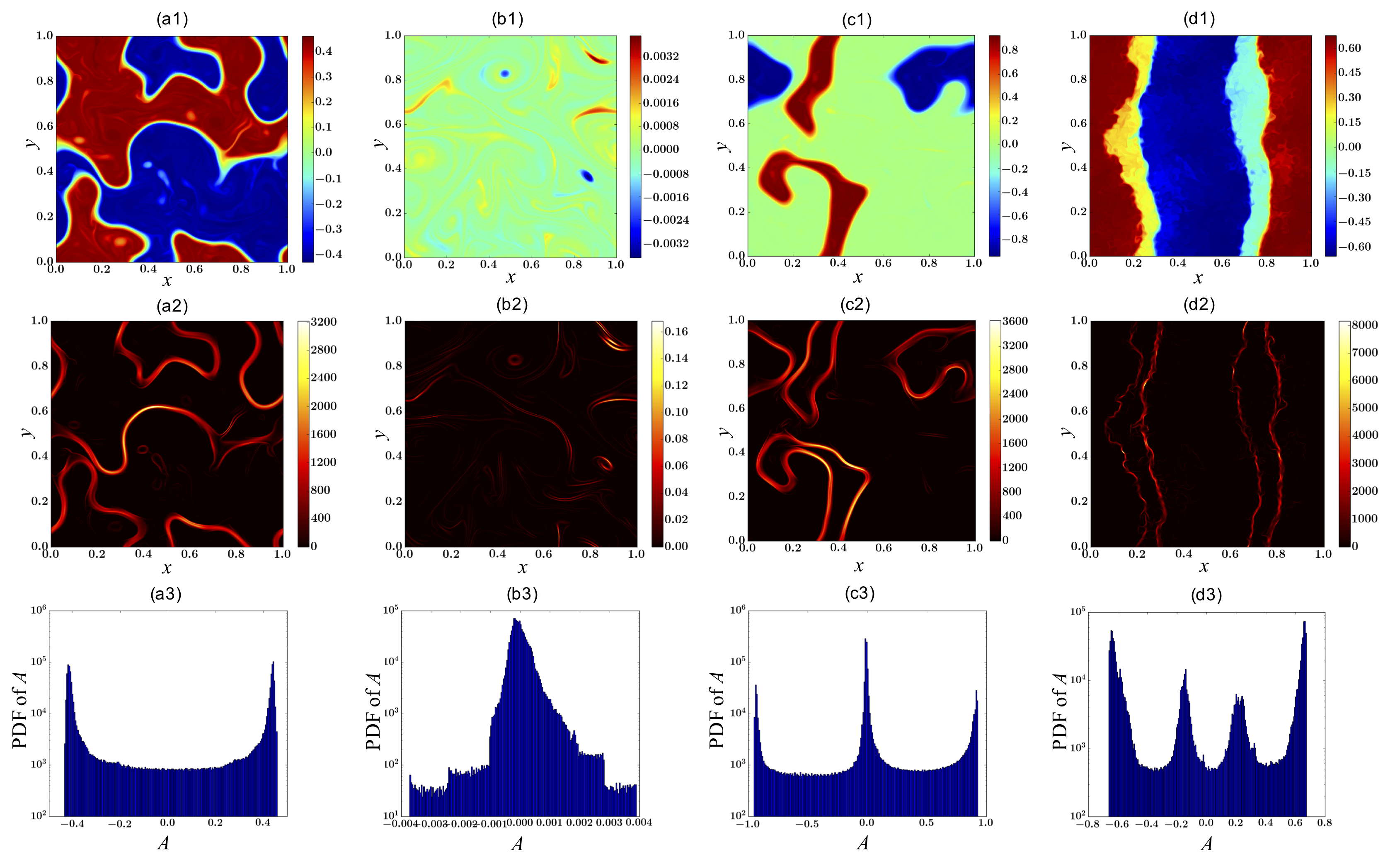}
    \caption{Row 1: $A$ field snapshots; Row 2: $B^2$ field snapshots; Row 3: PDF of $A$. Column a: Run1 at $t=10$ (suppression stage). The system exhibits blob-and-barrier feature, and the PDF of $A$ is bimodal. Column b: Run1 at $t=17$ (kinematic decay stage). The distribution of the fields are trivial. Column c: Run2 at $t=10$. Two peaks still arise on the PDF of $A$ even though its initial condition is unimodal. Column d: Run3 at $t=4.5$. The system exhibits staircases feature, and the PDF of $A$ has multiple peaks.}
    \label{snapshots}
\end{figure*}

Non-trivial blob-and-barrier structure is observed in real space at large $\mathrm{Rm}$, and this structure forms quickly after a short transition period. Fig.~\ref{snapshots} (a1) shows a snapshot of magnetic potential in the suppression stage for Run1. It consists of ``blobs'' (regions in red and blue) and interstices (green), and is very different from the initial condition, for which a mean field is relevant. Fig.~\ref{snapshots} (a2) shows the $B^2$ field for the same run. The high $B^2$ regions (bright color) occur at the interstices of the $A$ blobs, since $\mathbf{B}\equiv\hat{\mathbf{z}}\times\nabla A$. The interstices have a 1-dimensional shape. We call these 1-dimensional, high $B^2$ regions ``barriers'', because these are the regions where transport is strongly suppressed relative to the kinematic case $\eta_K$, due to locally strong  $B^2$, as discussed below. One measure of this blob-and-barrier structure is the structure of the probability density function (PDF) of $A$. As is shown in Fig.~\ref{snapshots} (a3), the PDF of $A$ for Run1 during the suppression stage has two peaks, both at $A\neq 0$. 

Notably, such a structure of the PDF also appears in the analogous CHNS system. Some binary fluid transfers from miscible phase to immiscible phase when the temperature dropped to below the corresponding critical temperature, and this second order phase transition is called spinodal decomposition. The Cahn-Hilliard Navier-Stokes (CHNS) equations describe a binary fluid undergoing spinodal decomposition:

\begin{align}
\partial_t\psi+\mathbf{v}\cdot\nabla\psi&=D\nabla^2(-\psi+\psi^3-\xi^2\nabla^2\psi) \label{CHNS1_2019}\\
\partial_t\omega+\mathbf{v}\cdot\nabla\omega&=\frac{\xi^2}{\rho}\mathbf{B}_\psi\cdot\nabla\nabla^2\psi+\nu\nabla^2\omega \label{CHNS2_2019}\\
\mathbf{v}=\mathbf{\hat{z}}\times\nabla\phi&,\ \omega = \nabla^2\phi \label{CHNS3_2019}\\
\mathbf{B}_\psi=\mathbf{\hat{z}}\times\nabla\psi&,\ j_\psi = \xi^2\nabla^2\psi \label{CHNS4_2019}
\end{align}
where $\psi = \frac{\rho_A-\rho_B}{\rho_A+\rho_B}$ is the local relative concentration, and $\xi$ is a coefficient describing the strength of the surface tension interaction.

\begin{figure}[htbp] 
    \centering
    \includegraphics[width=\columnwidth]{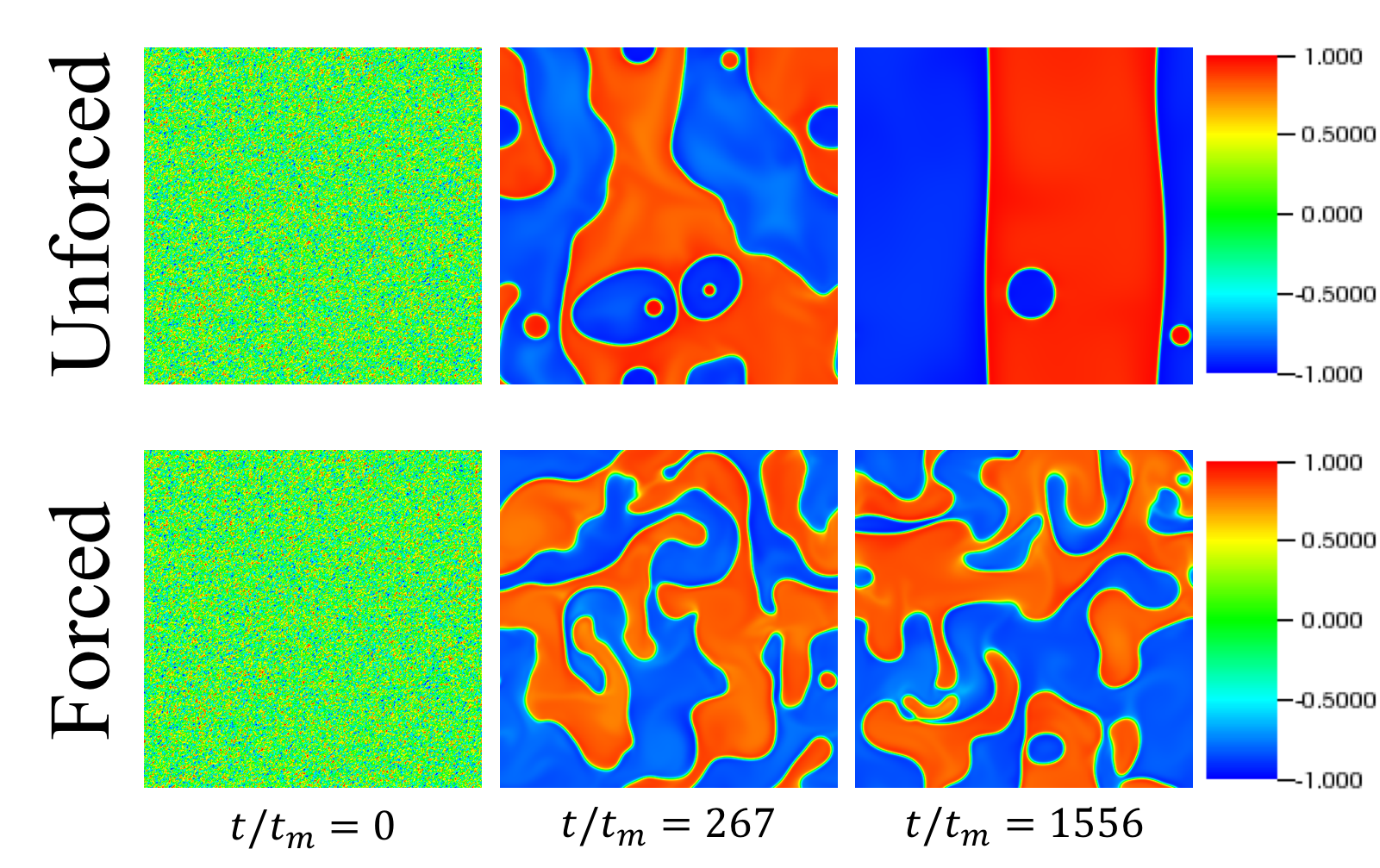}
    \caption{Some typical screenshots for the $\psi$ field in the 2D CHNS system. Reprint from \cite{fan_cascades_2016}.}
    \label{pcolor_2019}
\end{figure}

\begin{table}
\caption{The correspondence between 2D MHD and the 2D CHNS system. Reprint from \cite{fan_cascades_2016}.}
\begin{center}
\begin{tabular}{ccc}
\hline
\hline
& 2D MHD & 2D CHNS \\
\hline
Magnetic Potential & $A$ & $\psi$ \\
Magnetic Field & $\mathbf{B}$ & $\mathbf{B}_\psi$ \\
Current & $j$ & $j_\psi$ \\
Diffusivity & $\eta$ & $D$ \\
Interaction strength & $\frac{1}{\mu_0}$ & $\xi^2$ \\
\hline
\hline
\end{tabular}
\end{center}
\label{correspondence_2019}
\end{table}%

2D CHNS and 2D MHD are both active scalar systems. The two systems are analogous, and the correspondence of the physics quantities between the two systems are summarized in Table.~\ref{correspondence_2019}. The comparison and contrast of some most important features are summarized in Table.~\ref{comparison_2019} and Table.~\ref{contrast_2019}. See Ref.~\cite{fan_cascades_2016,fan_formation_2017,fan_chns:_2018} for more details about turbulence in 2D CHNS.

\begin{table*}
\caption{Comparison of 2D MHD and the 2D CHNS system. Reprint from \cite{fan_cascades_2016}.}
\scriptsize
\begin{center}
\begin{tabular}{ccc}
\hline
\hline
& 2D MHD & 2D CHNS\\
\hline
Ideal Quadratic Conserved Quantities & Conservation of $E$, $H^A$ and $H^C$ & Conservation of $E$, $H^\psi$ and $H^C$\\
Role of elastic waves & Alfven wave couples $\mathbf{v}$ with $\mathbf{B}$ & CHNS linear elastic wave couples $\mathbf{v}$ with $\mathbf{B}_\psi$\\
Origin of elasticity & Magnetic field induces elasticity & Surface tension induces elasticity\\
Origin of the inverse cascades & The coalescence of magnetic flux blobs & The coalescence of blobs of the same species\\
The inverse cascades & Inverse cascade of $H^A$ & Inverse cascade of $H^\psi$\\
Power law of spectra & $H^A_k\sim k^{-7/3}$ & $H^\psi_k\sim k^{-7/3}$\\
\hline
\hline
\end{tabular}
\end{center}
\label{comparison_2019}
\end{table*}

\begin{table*}
\caption{Contrast of 2D MHD and the 2D CHNS system. Reprint from \cite{fan_cascades_2016}.}
\scriptsize
\begin{center}
\begin{tabular}{ccc}
\hline
\hline
& 2D MHD & 2D CHNS\\
\hline
Diffusion & A simple positive diffusion term & A negative, a self nonlinear, and a hyper-diffusion term\\
Range of potential & No restriction for range of $A$ & $\psi\in[-1,1]$ \\
Interface Packing Fraction & Not far from $50\%$ & Small\\
Back reaction & $\mathbf{j}\times\mathbf{B}$ force can be significant & Back reaction is apparently limited\\
Kinetic energy spectrum & $E^K_k\sim k^{-3/2}$ & $E^K_k\sim k^{-3}$\\
Suggestive cascade by $E^K_k$ & Suggestive of direct energy cascade & Suggestive of direct enstrophy cascade\\
\hline
\hline
\end{tabular}
\end{center}
\label{contrast_2019}
\end{table*}

In comparison with the blob-and-barrier structure described above, in the kinematic decay stage of Run1 (i.e. at later time, when the magnetic field is so weak that $\eta_T$ reverts to $\eta_K$), the fields are well mixed and nontrivial real space structure is absent. No barriers are discernible in the decay stage. The corresponding PDF of $A$ is a distribution for a passive scalar, with \textit{one peak} at $A=0$, as shown in Fig.~\ref{snapshots} column (b).

\begin{figure*}[htbp] 
    \centering
    \includegraphics[width=\textwidth]{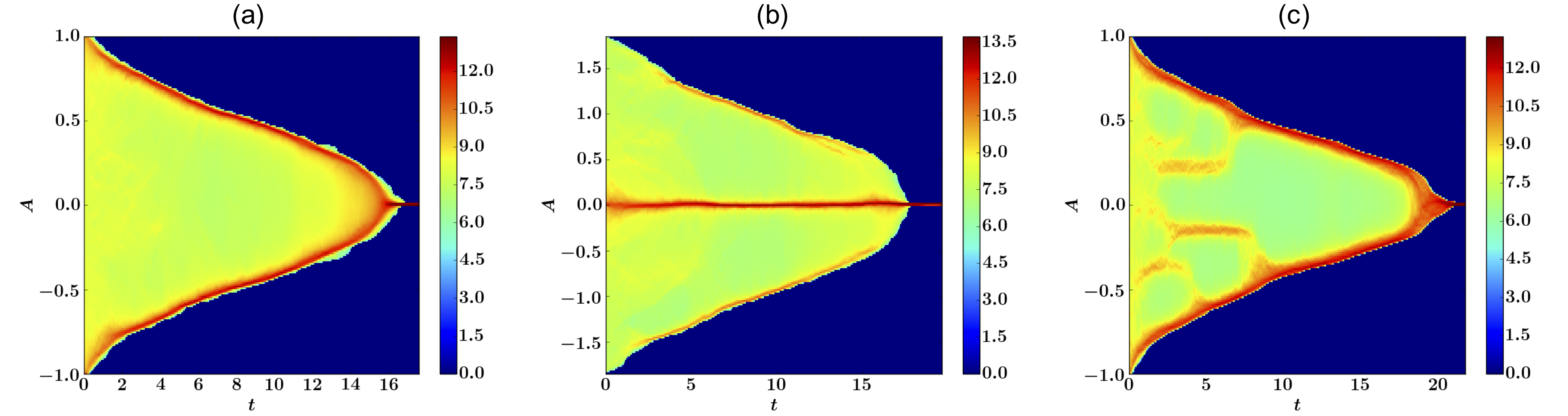}
    \caption{The time evolutions of PDF of $A$, and the values are in logarithm scale (base 10). (a) For Run1, the PDF is bimodal in the suppression stage, and $\Delta A$ between the two peaks reduces in time. The PDF becomes unimodal in the kinematic decay stage. (b) For Run2, two peaks at $A\neq 0$ still arise spontaneously given a unimodal initial condition. (c) For Run3, with external forcing at smaller scale, layering and coarsening can occur. See further explanations in the text.}
    \label{pdf_evolution}
\end{figure*}

The time evolution of PDF of $A$ for Run1 (Fig.~\ref{pdf_evolution} (a)) has a horizontal ``Y'' shape. The PDF has two peaks initially, and the interval between the peaks decreases as the $A$ field decays. The PDF changes from double peak to single peak as the system evolves from the suppression stage to the kinematic stage.

\begin{figure}[htbp] 
    \centering
    \includegraphics[width=\columnwidth]{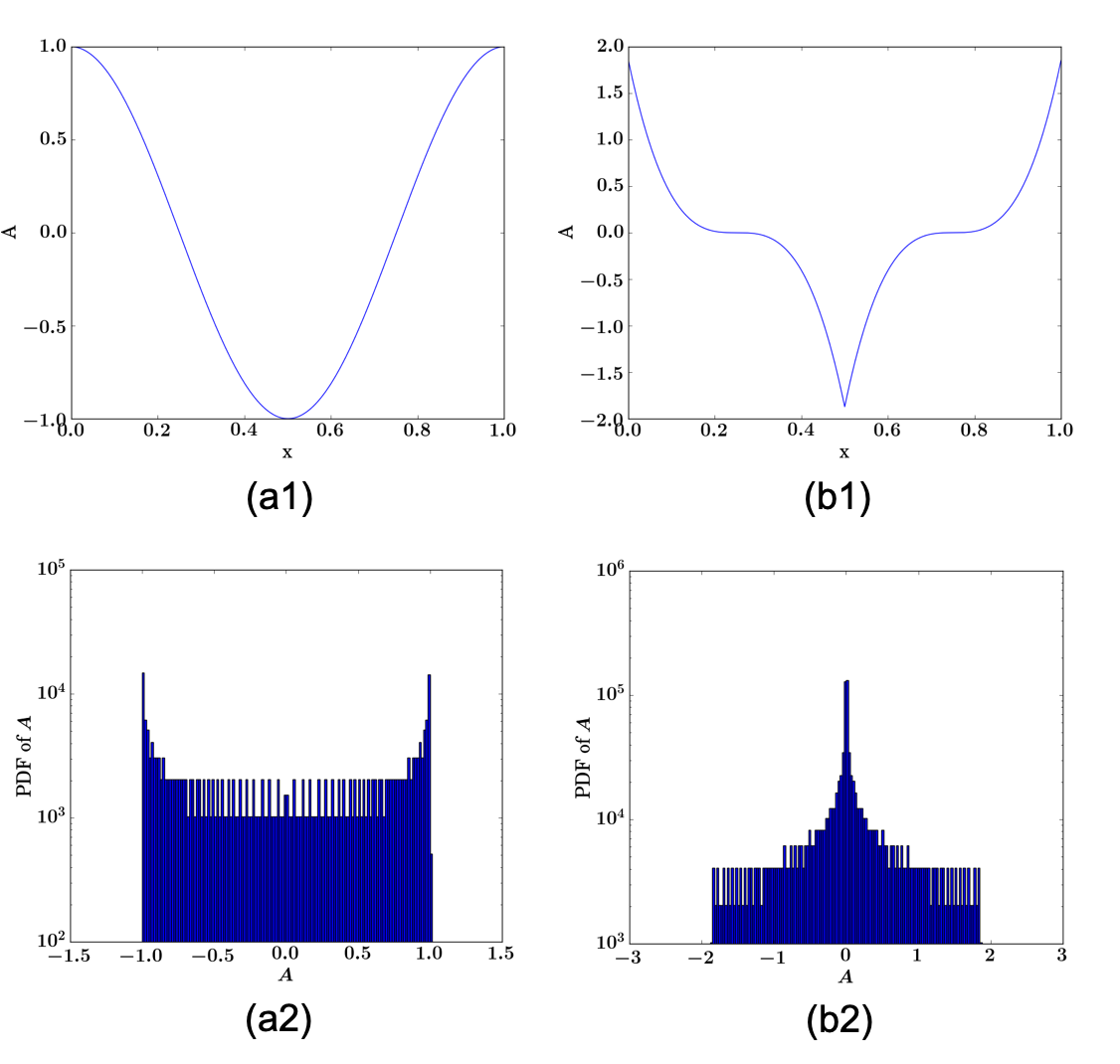}
    \caption{The initial conditions for $A$ and their PDFs: (a) ``Bimodal'' for Run1 and Run3; (b) ``Unimodal'' for Run2.}
    \label{initial_condition_fig}
\end{figure}

\begin{figure}[htbp] 
    \centering
    \includegraphics[width=\columnwidth]{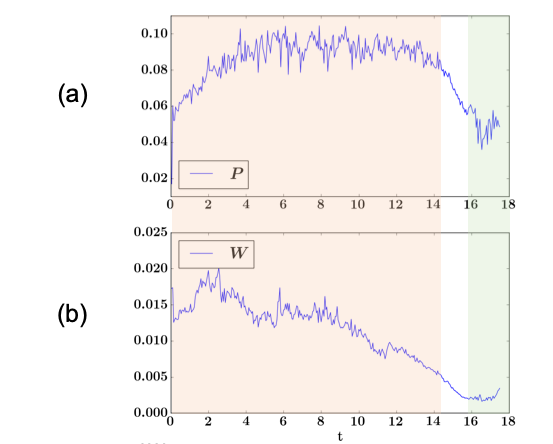}
    \caption{Time evolution of (a) packing fraction $P$; and (b) barrier width $W$ in Run1.}
    \label{P_W_time}
\end{figure}

Two quantities which characterize the field structure in the suppression stage are the \textit{packing fraction $P$}, and \textit{barrier width $W$}, defined below. In order to identify the barriers, we set a threshold on local field intensity, and define the barriers to be the regions where $B(x,y) > \sqrt{\langle B^2\rangle}*2$. The packing fraction $P$ is defined as:
\be 
P\equiv \frac{\text{\# of grid points in barrier regions}}{\text{\# of total grid points}}
\ee 
$P$ is the fraction of the space where intensity exceeds the mean square value. The expression for the barrier width is $W\sim\Delta A/B_b$, where $\Delta A$ is the difference in $A$ between adjacent blobs, and $B_b$ is the magnitude of the magnetic field in the barrier regions. We use $\sqrt{\langle A^2\rangle}$ to estimate $\Delta A$ for the bimodal PDF, such as for Run1. The narrow barriers contain most of the magnetic energy. For example, in Run1 at $t=10$, the barriers occupy only $P=9.9\%$ of the system space, but the magnetic field in these regions accounts for $80.7\%$ of the magnetic energy. Therefore, we can use the magnetic energy in the barriers $\langle B_b^2\rangle$ to approximate the total magnetic energy, i.e.:
\be
\sum_{\text{barriers}}B_b^2\sim\int\intd^2 xB^2
\ee
It follows that $\langle B_b^2\rangle \sim \langle B^2\rangle /P$. We can thus define $W$ based on the arguments above as:
\be 
W^2\equiv \langle A^2\rangle/(\langle B^2\rangle/P)
\ee 
This definition of $W$ can be justified by measuring the approximate barrier widths. The time evolutions of $P$ and $W$ in Run1 are shown in Fig.~\ref{P_W_time}. $P$ stays at $0.08\sim0.10$ throughout the suppression stage. $P$ starts to decline near the end of the suppression stage, and drops to the noise level in the kinematic decay stage. $W$ decreases during the suppression stage, due mainly to the decrease in $\Delta A$. It is important to note that the decline in $P$, which begins at $t\sim 13$, slightly \textit{leads} the decay in magnetic energy, which begins at $t\sim 15$. This supports the notion that barriers, the population of which is measured by $P$, are responsible for the quenching of mixing and decay in the suppression stage.

One may question whether the bimodal PDF is due to the initial condition, since the cosine initial condition in Run1 is bimodal. The answer is no. In order to show this, a unimodal initial condition is constructed for Run2, such that the initial PDF of $A$ has one peak at $A=0$:
\be
A_I(x,y) = A_0*
\begin{cases} 
      -(x-0.25)^3 & 0 <= x < 1/2 \\
      (x-0.75)^3 & 1/2 <= x < 1 
   \end{cases}
\ee

See Fig.~\ref{initial_condition_fig} for the comparison between bimodal and unimodal initial condition. To make Run2 and Run1 have the same time duration of the suppression stage, the initial magnitude $A_0$ in Run2 is chosen such that the initial $\langle A^2\rangle$ (not $E_B$!) is the same with Run1.

Fig.~\ref{snapshots} column (c) shows a snapshot for Run2 at $t=10$. The time evolution of the PDF of $A$ for that case is shown in Fig.~\ref{pdf_evolution} (b). It is evident that, two non zero peaks in the PDF of $A$ still arise, even if the initial condition is unimodal. The blob structure in $A$ and the barrier structure in $B^2$ are also evident.

\section{Analysis: local}

One can easily see from the $B^2$ fields plots in Fig.~\ref{snapshots} that, a large scale $\langle\mathbf{B}\rangle$ does not exist. Intermittent magnetic intensity, with low $P$, is a consequence of the blob-and-barrier structure. Therefore, the traditional approach of mean field theory, especially Eqn.~(\ref{b_avg_eqn}), is neither applicable nor relevant. Globally, no theory exists for $\mathbf{B}_0=0$. Usual closure approaches appear useful when the averaging window is restricted to a suitable size, corresponding to a \textit{localized region} within which a mean $\mathbf{B}$ exists. In order to derive an expression for the effective $\eta_T$ for such a local region from dynamics, we extend the theory by \cite{gruzinov_self-consistent_1994,gruzinov_nonlinear_1996,diamond_self-consistent_2005,diamond_modern_2010}, and propose:
\be 
\eta_T=\frac{\langle v^2\rangle^{1/2} l}{1+\mathrm{Rm}\frac{1}{\mu_0\rho}\langle\mathbf{B}\rangle^2/\langle v^2\rangle+\mathrm{Rm}\frac{1}{\mu_0\rho}\frac{\langle A^2\rangle}{L_{blob}^2}/\langle v^2\rangle}\label{eta_T_eqn_new}
\ee 
Here $L_{blob}$ is the size of the large $A$ blobs, i.e. the characteristic length scale for $\langle A^2\rangle$. The derivation is shown below. 

We start from:
\be 
\frac12[\partial_t \langle A^2\rangle+\langle\nabla\cdot(\mathbf{v} A^2)\rangle]=-\Gamma_A\frac{\partial\langle A\rangle}{\partial x}-\eta\langle B^2\rangle\label{several_terms}
\ee 
where $\Gamma_A=\langle v_x A\rangle$ is the spatial flux of $A$. In the past, only the $\Gamma_A\frac{\partial\langle A\rangle}{\partial x}$ term is kept in (\ref{several_terms}) to balance $\eta\langle B^2\rangle$. However, in the absence of $\mathbf{B}_0$, $\Gamma_A\frac{\partial\langle A\rangle}{\partial x}$ term can be small, while the triplet term $\langle\nabla\cdot(\mathbf{v}A^2)\rangle$ can remain, if the average is taken over a window smaller than the system size $L_0$. Note the relevant scale $l$ here is
\be 
l_d<W<l<L_0
\ee
where $l_d$ is the dissipation scale. Retaining all contributions, we have
\be 
\partial_t\langle A^2\rangle=-\langle\mathbf{v}A\rangle\cdot\nabla \langle A\rangle-\nabla\cdot\langle\mathbf{v} A^2\rangle-\eta\langle B^2\rangle
\ee 
Now assume the fluxes are Fickian. Note that, in principle, there are two diffusion coefficients:
\begin{align}
\langle\mathbf{v}A\rangle &= -\eta_{T1}\nabla\langle A\rangle\\
\langle\mathbf{v}A^2\rangle &= -\eta_{T2}\nabla\langle A^2\rangle
\end{align}
Plugging them in, we get
\be 
\partial_t\langle A^2\rangle=\eta_{T1}(\nabla\langle A\rangle)^2+\nabla\eta_{T2}\cdot\nabla\langle A^2\rangle-\eta\langle B^2\rangle
\ee 
The first term on the RHS is turbulent diffusion of $\langle A\rangle$, corresponding to the large scale magnetic field. The second term is the turbulent diffusion of $\langle A^2\rangle$, which controls decay in weak magnetic field. The third term is the usual collisional dissipation. In principle, $\eta_{T1}\neq\eta_{T2}$, though these two are related. Both terms are retained. For simplicity, we assume $\eta_{T1}=\eta_{T2}=\eta_T$. For a stationary state, we have
\be 
\langle B^2\rangle\sim\frac{\eta_T}{\eta}(\langle B\rangle^2+\langle A^2\rangle/L_{blob}^2)\label{new_b2}
\ee 
where $L_{blob}$ is the blob size, the characteristic length scale for $\langle A^2\rangle$. By standard closure methods, one can obtain an expression for $\eta_T$ \cite{pouquet_strong_1976,pouquet_two-dimensional_1978}:
\be 
\eta_T=\sum_{\mathbf{k}}\tau_c[\langle v^2\rangle_\mathbf{k}-\frac{1}{\mu_0\rho}\langle B^2\rangle_\mathbf{k}]\label{eta_T_v2_minus_b2}
\ee 
Plugging (\ref{new_b2}) into (\ref{eta_T_v2_minus_b2}) yields Eqn.~(\ref{eta_T_eqn_new}) proposed above. Detailed comparisons of Eqn.~(\ref{eta_T_eqn_new}) with simulation results are nontrivial and will be left for a future paper.

Note that $L_{blob}\ll L_0$. In regions where no high intensity magnetic field is present, i.e. inside blobs, $\mathrm{Rm}\frac{1}{\mu_0\rho}\langle B\rangle^2/\langle v^2\rangle$ is negligible. Yet transport is still reduced relative to kinematics by $\langle A^2\rangle$, via the $\mathrm{Rm}\frac{1}{\mu_0\rho}\frac{\langle A^2\rangle}{L_{blob}^2}/\langle v^2\rangle$ term. In the barrier regions where magnetic energy is large, $\mathrm{Rm}\frac{1}{\mu_0\rho}\langle B\rangle^2/\langle v^2\rangle$ is dominant, since $\langle B^2\rangle\gg\langle A^2\rangle/L^2_{blob}$ for $P\ll 1$. Such regions -- barriers -- are where turbulent transport of $A$ is most strongly suppressed.

A key question concerns how transport barriers form spontaneously in turbulent 2D MHD. We argue that transport barriers result from negative resistivity, driven by the inverse cascade of $\langle A^2\rangle$. In Eqn.~(\ref{eta_T_v2_minus_b2}), the positive contribution to $\eta_T$ is a consequence of turbulent mixing by fluid advection, while the second, negative, term is a consequence of flux coalescence. From the above, we see that the turbulent resistivity can go negative locally, where $\langle B^2\rangle$ is strong. Of course, the system-averaged resistivity is positive, so the field decays, though slowly. Note though that a local negative contribution can trigger a feedback loop, i.e.: $B^2$ strong in a specific region $\rightarrow$ local $\eta_T$ negative $\rightarrow$ local $\nabla A$ increases $\rightarrow$ local $B^2$ increases further. The feedback process saturates after the short transition period, as the inverse cascade of $\langle A^2\rangle$ must ultimately deplete the small scales.

\begin{figure}[htbp] 
    \centering
    \includegraphics[width=0.9\columnwidth]{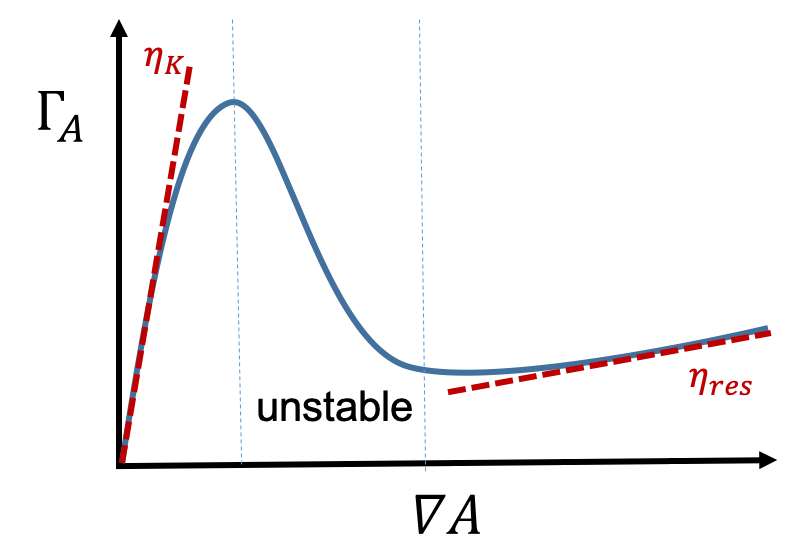}
    \caption{A sketch showing the relationship between flux $\Gamma_A$ and $\nabla A$. The total resistivity is $\eta_{tot}=\delta\Gamma_A/\delta\nabla A$, and is composed of turbulent and collisional parts $\eta_{tot}=\eta_T+\eta$. In the small B limit, $\eta_{tot}\sim\eta_K$; in the large B limit, the residual resistivity is $\eta_{res}\sim\frac{ul}{\mathrm{Rm}\frac{1}{\mu_0\rho}\langle B\rangle^2/\langle v^2\rangle+\mathrm{Rm}\frac{1}{\mu_0\rho}\frac{\langle A^2\rangle}{L_{blob}^2}/\langle v^2\rangle}+\eta$. The transition between the two limits is the transport bifurcation.}
    \label{the_S_curve}
\end{figure}

Another way to view this evolution is as a local transport bifurcation -- see the spatially local S-shaped flux-gradient curve for $A$, shown in Fig.~\ref{the_S_curve} for illustration, which follows from Eqn.~(\ref{eta_T_eqn_new}). The S-curve describes a bi-stable system. Note there are two stable ranges with positive slope, and one unstable region between, with a negative slope (as for negative resistivity). This implies that barrier formation is a transport bifurcation, which occurs when local magnetic intensity exceeds the threshold given by (\ref{eta_T_v2_minus_b2}). This mechanism resembles a transport bifurcation in magnetically confined systems
\cite{ashourvan_how_2016,ashourvan_emergence_2017}. Here, feedback via regions of locally intense $B^2$, rather than $\mathbf{E}\times\mathbf{B}$ shear, is the trigger for barrier formation.

\section{Layering of magnetic potential}

Inhomogeneous mixing and bistability (of which negative viscosity is a symptom) are the key elements in the dynamics of layering (i.e. staircase formation) in many systems \cite{balmforth_dynamics_1998,ashourvan_how_2016,ashourvan_emergence_2017}. Given that, and the ubiquitous blob-and-barrier structure here, it is natural to ask if spontaneous layering can occur in 2D MHD. We answer in the affirmative -- see Fig.~\ref{snapshots} column (d). The initial condition in Run3 is the same bimodal one as for Run1. The key difference in parameters is the forcing scale, which is smaller here, i.e. $k=32$ in Run3, rather than $k=5$ for the other runs. As shown in Fig.~\ref{pdf_evolution} (c), the layered structure consists of regions of homogenized $A$, with small transition layers of sharp gradients in $A$ (and thus $B^2$) between them. Layering thus induces transport barriers. The layered structure persists for the duration of the suppression stage, but coarsens, as shown in Fig.~\ref{pdf_evolution} (c). Coarsening occurs by a sequence of blob mergers. Note that by $t\sim 4$ the staircase has coarsened to four transition layers. We note that closure theory for the evolution of $\langle A^2\rangle_\mathbf{k}$ predict a positive turbulent hyper-resistivity along with the negative component of the turbulent resistivity \cite{diamond_self-consistent_2005,diamond_modern_2010}. This implies that evolution equation for mean $\langle A\rangle$ has a structure much like the Cahn-Hilliard equation, the solutions of which are known to manifest mergers and coarsening \cite{fan_chns:_2018}.

\section{Conclusions}

In summary, we observe a blob-and-barrier real space structure in the decay of magnetic fields in turbulent 2D MHD. The magnetic field and the resulting barriers are highly intermittent, and cannot be treated by mean field theory. The turbulent resistivity is suppressed in the barriers, where the magnetic fields are strong. The barriers form at blob interstices. For small scale forcing, spontaneous layering of magnetic potential occurs due to inhomogeneous mixing. Barriers form between layers. The layered structure coarsens in time. 

This analysis has implications beyond 2D. One line of development is to the quenching of transport of magnetic helicity and magnetic dynamo processes by spatially intermittent but locally strong magnetic fields. The other is to anomalous dissipation in anisotropically ordered 3D systems, such as reduced MHD, where the nonlinear dynamics are effectively two dimensional. These topics will be pursued in the future.

\begin{acknowledgments}
This research was supported by the US Department of Energy, Office of Science, Office of Fusion Energy Sciences, under Award No. DE-FG02-04ER54738 and CMTFO Award No. DE-SC0008378.
\end{acknowledgments}

\bibliography{MHD_transport} 

\begin{thebibliography}{40}%
\makeatletter
\providecommand \@ifxundefined [1]{%
 \@ifx{#1\undefined}
}%
\providecommand \@ifnum [1]{%
 \ifnum #1\expandafter \@firstoftwo
 \else \expandafter \@secondoftwo
 \fi
}%
\providecommand \@ifx [1]{%
 \ifx #1\expandafter \@firstoftwo
 \else \expandafter \@secondoftwo
 \fi
}%
\providecommand \natexlab [1]{#1}%
\providecommand \enquote  [1]{``#1''}%
\providecommand \bibnamefont  [1]{#1}%
\providecommand \bibfnamefont [1]{#1}%
\providecommand \citenamefont [1]{#1}%
\providecommand \href@noop [0]{\@secondoftwo}%
\providecommand \href [0]{\begingroup \@sanitize@url \@href}%
\providecommand \@href[1]{\@@startlink{#1}\@@href}%
\providecommand \@@href[1]{\endgroup#1\@@endlink}%
\providecommand \@sanitize@url [0]{\catcode `\\12\catcode `\$12\catcode
  `\&12\catcode `\#12\catcode `\^12\catcode `\_12\catcode `\%12\relax}%
\providecommand \@@startlink[1]{}%
\providecommand \@@endlink[0]{}%
\providecommand \url  [0]{\begingroup\@sanitize@url \@url }%
\providecommand \@url [1]{\endgroup\@href {#1}{\urlprefix }}%
\providecommand \urlprefix  [0]{URL }%
\providecommand \Eprint [0]{\href }%
\providecommand \doibase [0]{http://dx.doi.org/}%
\providecommand \selectlanguage [0]{\@gobble}%
\providecommand \bibinfo  [0]{\@secondoftwo}%
\providecommand \bibfield  [0]{\@secondoftwo}%
\providecommand \translation [1]{[#1]}%
\providecommand \BibitemOpen [0]{}%
\providecommand \bibitemStop [0]{}%
\providecommand \bibitemNoStop [0]{.\EOS\space}%
\providecommand \EOS [0]{\spacefactor3000\relax}%
\providecommand \BibitemShut  [1]{\csname bibitem#1\endcsname}%
\let\auto@bib@innerbib\@empty
\bibitem [{\citenamefont {Moffatt}(1983)}]{moffatt_magnetic_1983}%
  \BibitemOpen
  \bibfield  {author} {\bibinfo {author} {\bibfnamefont {H.~K.}\ \bibnamefont
  {Moffatt}},\ }\href@noop {} {\emph {\bibinfo {title} {Magnetic field
  generation in electrically conducting fluids}}}\ (\bibinfo  {publisher}
  {Cambridge University Press},\ \bibinfo {year} {1983})\BibitemShut {NoStop}%
\bibitem [{\citenamefont {Cattaneo}\ and\ \citenamefont
  {Vainshtein}(1991)}]{cattaneo_suppression_1991}%
  \BibitemOpen
  \bibfield  {author} {\bibinfo {author} {\bibfnamefont {F.}~\bibnamefont
  {Cattaneo}}\ and\ \bibinfo {author} {\bibfnamefont {S.}~\bibnamefont
  {Vainshtein}},\ }\href {http://adsabs.harvard.edu/full/1991ApJ...376L..21C}
  {\bibfield  {journal} {\bibinfo  {journal} {The Astrophysical Journal}\ }
  (\bibinfo {year} {1991})}\BibitemShut {NoStop}%
\bibitem [{\citenamefont {Vainshtein}\ and\ \citenamefont
  {Rosner}(1991)}]{vainshtein_turbulent_1991}%
  \BibitemOpen
  \bibfield  {author} {\bibinfo {author} {\bibfnamefont {S.~I.}\ \bibnamefont
  {Vainshtein}}\ and\ \bibinfo {author} {\bibfnamefont {R.}~\bibnamefont
  {Rosner}},\ }\href {\doibase 10.1086/170268} {\bibfield  {journal} {\bibinfo
  {journal} {The Astrophysical Journal}\ }\textbf {\bibinfo {volume} {376}},\
  \bibinfo {pages} {199} (\bibinfo {year} {1991})}\BibitemShut {NoStop}%
\bibitem [{\citenamefont {Vainshtein}\ and\ \citenamefont
  {Cattaneo}(1992)}]{vainshtein_nonlinear_1992}%
  \BibitemOpen
  \bibfield  {author} {\bibinfo {author} {\bibfnamefont {S.}~\bibnamefont
  {Vainshtein}}\ and\ \bibinfo {author} {\bibfnamefont {F.}~\bibnamefont
  {Cattaneo}},\ }\href
  {http://articles.adsabs.harvard.edu/full/1992ApJ...393..165V} {\bibfield
  {journal} {\bibinfo  {journal} {The Astrophysical Journal}\ } (\bibinfo
  {year} {1992})}\BibitemShut {NoStop}%
\bibitem [{\citenamefont {Cattaneo}(1994)}]{cattaneo_effects_1994}%
  \BibitemOpen
  \bibfield  {author} {\bibinfo {author} {\bibfnamefont {F.}~\bibnamefont
  {Cattaneo}},\ }\href {http://adsabs.harvard.edu/full/1994ApJ...434..200C}
  {\bibfield  {journal} {\bibinfo  {journal} {The Astrophysical Journal}\
  }\textbf {\bibinfo {volume} {434}},\ \bibinfo {pages} {200} (\bibinfo {year}
  {1994})}\BibitemShut {NoStop}%
\bibitem [{\citenamefont {Balmforth}\ \emph {et~al.}(1998)\citenamefont
  {Balmforth}, \citenamefont {Smith},\ and\ \citenamefont
  {Young}}]{balmforth_dynamics_1998}%
  \BibitemOpen
  \bibfield  {author} {\bibinfo {author} {\bibfnamefont {N.~J.}\ \bibnamefont
  {Balmforth}}, \bibinfo {author} {\bibfnamefont {S.~G.~L.}\ \bibnamefont
  {Smith}}, \ and\ \bibinfo {author} {\bibfnamefont {W.~R.}\ \bibnamefont
  {Young}},\ }\href {\doibase 10.1017/S0022112097007970} {\bibfield  {journal}
  {\bibinfo  {journal} {Journal of Fluid Mechanics}\ }\textbf {\bibinfo
  {volume} {355}},\ \bibinfo {pages} {329} (\bibinfo {year}
  {1998})}\BibitemShut {NoStop}%
\bibitem [{\citenamefont {Lazarian}\ and\ \citenamefont
  {Vishniac}(1999)}]{lazarian_reconnection_1999}%
  \BibitemOpen
  \bibfield  {author} {\bibinfo {author} {\bibfnamefont {A.}~\bibnamefont
  {Lazarian}}\ and\ \bibinfo {author} {\bibfnamefont {E.~T.}\ \bibnamefont
  {Vishniac}},\ }\href {\doibase 10.1086/307233} {\bibfield  {journal}
  {\bibinfo  {journal} {The Astrophysical Journal}\ }\textbf {\bibinfo {volume}
  {517}},\ \bibinfo {pages} {700} (\bibinfo {year} {1999})}\BibitemShut
  {NoStop}%
\bibitem [{\citenamefont {Biskamp}\ and\ \citenamefont
  {Schwarz}(2001)}]{biskamp_two-dimensional_2001}%
  \BibitemOpen
  \bibfield  {author} {\bibinfo {author} {\bibfnamefont {D.}~\bibnamefont
  {Biskamp}}\ and\ \bibinfo {author} {\bibfnamefont {E.}~\bibnamefont
  {Schwarz}},\ }\href {\doibase 10.1063/1.1377611} {\bibfield  {journal}
  {\bibinfo  {journal} {Physics of Plasmas}\ }\textbf {\bibinfo {volume} {8}},\
  \bibinfo {pages} {3282} (\bibinfo {year} {2001})}\BibitemShut {NoStop}%
\bibitem [{\citenamefont {Field}\ and\ \citenamefont
  {Blackman}(2002)}]{field_dynamical_2002}%
  \BibitemOpen
  \bibfield  {author} {\bibinfo {author} {\bibfnamefont {G.~B.}\ \bibnamefont
  {Field}}\ and\ \bibinfo {author} {\bibfnamefont {E.~G.}\ \bibnamefont
  {Blackman}},\ }\href {\doibase 10.1086/340233} {\bibfield  {journal}
  {\bibinfo  {journal} {The Astrophysical Journal}\ }\textbf {\bibinfo {volume}
  {572}},\ \bibinfo {pages} {685} (\bibinfo {year} {2002})}\BibitemShut
  {NoStop}%
\bibitem [{\citenamefont {Mininni}\ \emph {et~al.}(2005)\citenamefont
  {Mininni}, \citenamefont {Montgomery},\ and\ \citenamefont
  {Pouquet}}]{mininni_numerical_2005}%
  \BibitemOpen
  \bibfield  {author} {\bibinfo {author} {\bibfnamefont {P.~D.}\ \bibnamefont
  {Mininni}}, \bibinfo {author} {\bibfnamefont {D.~C.}\ \bibnamefont
  {Montgomery}}, \ and\ \bibinfo {author} {\bibfnamefont {A.~G.}\ \bibnamefont
  {Pouquet}},\ }\href {\doibase 10.1063/1.1863260} {\bibfield  {journal}
  {\bibinfo  {journal} {Physics of Fluids}\ }\textbf {\bibinfo {volume} {17}},\
  \bibinfo {pages} {035112} (\bibinfo {year} {2005})}\BibitemShut {NoStop}%
\bibitem [{\citenamefont {Kim}(2006)}]{kim_consistent_2006}%
  \BibitemOpen
  \bibfield  {author} {\bibinfo {author} {\bibfnamefont {E.-j.}\ \bibnamefont
  {Kim}},\ }\href {\doibase 10.1103/PhysRevLett.96.084504} {\bibfield
  {journal} {\bibinfo  {journal} {Physical Review Letters}\ }\textbf {\bibinfo
  {volume} {96}},\ \bibinfo {pages} {084504} (\bibinfo {year}
  {2006})}\BibitemShut {NoStop}%
\bibitem [{\citenamefont {Silvers}(2005)}]{silvers_dynamic_2005}%
  \BibitemOpen
  \bibfield  {author} {\bibinfo {author} {\bibfnamefont {L.~J.}\ \bibnamefont
  {Silvers}},\ }\href {\doibase 10.1016/j.physleta.2004.11.043} {\bibfield
  {journal} {\bibinfo  {journal} {Physics Letters A}\ }\textbf {\bibinfo
  {volume} {334}},\ \bibinfo {pages} {400} (\bibinfo {year}
  {2005})}\BibitemShut {NoStop}%
\bibitem [{\citenamefont {Silvers}(2006)}]{silvers_choice_2006}%
  \BibitemOpen
  \bibfield  {author} {\bibinfo {author} {\bibfnamefont {L.~J.}\ \bibnamefont
  {Silvers}},\ }\href {\doibase 10.1111/j.1365-2966.2006.10008.x} {\bibfield
  {journal} {\bibinfo  {journal} {Monthly Notices of the Royal Astronomical
  Society}\ }\textbf {\bibinfo {volume} {367}},\ \bibinfo {pages} {1155}
  (\bibinfo {year} {2006})}\BibitemShut {NoStop}%
\bibitem [{\citenamefont {Kleeorin}\ and\ \citenamefont
  {Rogachevskii}(2007)}]{kleeorin_nonlinear_2007}%
  \BibitemOpen
  \bibfield  {author} {\bibinfo {author} {\bibfnamefont {N.}~\bibnamefont
  {Kleeorin}}\ and\ \bibinfo {author} {\bibfnamefont {I.}~\bibnamefont
  {Rogachevskii}},\ }\href {\doibase 10.1103/PhysRevE.75.066315} {\bibfield
  {journal} {\bibinfo  {journal} {Physical Review E}\ }\textbf {\bibinfo
  {volume} {75}},\ \bibinfo {pages} {066315} (\bibinfo {year}
  {2007})}\BibitemShut {NoStop}%
\bibitem [{\citenamefont {Keating}\ and\ \citenamefont
  {Diamond}(2007)}]{keating_turbulent_2007}%
  \BibitemOpen
  \bibfield  {author} {\bibinfo {author} {\bibfnamefont {S.~R.}\ \bibnamefont
  {Keating}}\ and\ \bibinfo {author} {\bibfnamefont {P.~H.}\ \bibnamefont
  {Diamond}},\ }\href {\doibase 10.1103/PhysRevLett.99.224502} {\bibfield
  {journal} {\bibinfo  {journal} {Physical Review Letters}\ }\textbf {\bibinfo
  {volume} {99}},\ \bibinfo {pages} {224502} (\bibinfo {year}
  {2007})}\BibitemShut {NoStop}%
\bibitem [{\citenamefont {Keating}\ \emph {et~al.}(2008)\citenamefont
  {Keating}, \citenamefont {Silvers},\ and\ \citenamefont
  {Diamond}}]{keating_cross-phase_2008}%
  \BibitemOpen
  \bibfield  {author} {\bibinfo {author} {\bibfnamefont {S.~R.}\ \bibnamefont
  {Keating}}, \bibinfo {author} {\bibfnamefont {L.~J.}\ \bibnamefont
  {Silvers}}, \ and\ \bibinfo {author} {\bibfnamefont {P.~H.}\ \bibnamefont
  {Diamond}},\ }\href {\doibase 10.1086/588654} {\bibfield  {journal} {\bibinfo
   {journal} {The Astrophysical Journal Letters}\ }\textbf {\bibinfo {volume}
  {678}},\ \bibinfo {pages} {L137} (\bibinfo {year} {2008})}\BibitemShut
  {NoStop}%
\bibitem [{\citenamefont {Keating}\ and\ \citenamefont
  {Diamond}(2008)}]{keating_turbulent_2008}%
  \BibitemOpen
  \bibfield  {author} {\bibinfo {author} {\bibfnamefont {S.~R.}\ \bibnamefont
  {Keating}}\ and\ \bibinfo {author} {\bibfnamefont {P.~H.}\ \bibnamefont
  {Diamond}},\ }\href {\doibase 10.1017/S002211200700941X} {\bibfield
  {journal} {\bibinfo  {journal} {Journal of Fluid Mechanics}\ }\textbf
  {\bibinfo {volume} {595}},\ \bibinfo {pages} {173} (\bibinfo {year}
  {2008})}\BibitemShut {NoStop}%
\bibitem [{\citenamefont {Tobias}\ \emph {et~al.}(2007)\citenamefont {Tobias},
  \citenamefont {Diamond},\ and\ \citenamefont {Hughes}}]{tobias_-plane_2007}%
  \BibitemOpen
  \bibfield  {author} {\bibinfo {author} {\bibfnamefont {S.~M.}\ \bibnamefont
  {Tobias}}, \bibinfo {author} {\bibfnamefont {P.~H.}\ \bibnamefont {Diamond}},
  \ and\ \bibinfo {author} {\bibfnamefont {D.~W.}\ \bibnamefont {Hughes}},\
  }\href {\doibase 10.1086/521978} {\bibfield  {journal} {\bibinfo  {journal}
  {The Astrophysical Journal Letters}\ }\textbf {\bibinfo {volume} {667}},\
  \bibinfo {pages} {L113} (\bibinfo {year} {2007})}\BibitemShut {NoStop}%
\bibitem [{\citenamefont {Eyink}\ \emph {et~al.}(2011)\citenamefont {Eyink},
  \citenamefont {Lazarian},\ and\ \citenamefont {Vishniac}}]{eyink_fast_2011}%
  \BibitemOpen
  \bibfield  {author} {\bibinfo {author} {\bibfnamefont {G.~L.}\ \bibnamefont
  {Eyink}}, \bibinfo {author} {\bibfnamefont {A.}~\bibnamefont {Lazarian}}, \
  and\ \bibinfo {author} {\bibfnamefont {E.~T.}\ \bibnamefont {Vishniac}},\
  }\href {\doibase 10.1088/0004-637X/743/1/51} {\bibfield  {journal} {\bibinfo
  {journal} {The Astrophysical Journal}\ }\textbf {\bibinfo {volume} {743}},\
  \bibinfo {pages} {51} (\bibinfo {year} {2011})}\BibitemShut {NoStop}%
\bibitem [{\citenamefont {Kondić}\ \emph {et~al.}(2016)\citenamefont
  {Kondić}, \citenamefont {Hughes},\ and\ \citenamefont
  {Tobias}}]{kondic_decay_2016}%
  \BibitemOpen
  \bibfield  {author} {\bibinfo {author} {\bibfnamefont {T.}~\bibnamefont
  {Kondić}}, \bibinfo {author} {\bibfnamefont {D.~W.}\ \bibnamefont {Hughes}},
  \ and\ \bibinfo {author} {\bibfnamefont {S.~M.}\ \bibnamefont {Tobias}},\
  }\href {\doibase 10.3847/0004-637X/823/2/111} {\bibfield  {journal} {\bibinfo
   {journal} {The Astrophysical Journal}\ }\textbf {\bibinfo {volume} {823}},\
  \bibinfo {pages} {111} (\bibinfo {year} {2016})}\BibitemShut {NoStop}%
\bibitem [{\citenamefont {Mak}\ \emph {et~al.}(2017)\citenamefont {Mak},
  \citenamefont {Griffiths},\ and\ \citenamefont {Hughes}}]{mak_vortex_2017}%
  \BibitemOpen
  \bibfield  {author} {\bibinfo {author} {\bibfnamefont {J.}~\bibnamefont
  {Mak}}, \bibinfo {author} {\bibfnamefont {S.~D.}\ \bibnamefont {Griffiths}},
  \ and\ \bibinfo {author} {\bibfnamefont {D.~W.}\ \bibnamefont {Hughes}},\
  }\href {\doibase 10.1103/PhysRevFluids.2.113701} {\bibfield  {journal}
  {\bibinfo  {journal} {Physical Review Fluids}\ }\textbf {\bibinfo {volume}
  {2}},\ \bibinfo {pages} {113701} (\bibinfo {year} {2017})}\BibitemShut
  {NoStop}%
\bibitem [{\citenamefont {Xi}\ \emph {et~al.}(2014{\natexlab{a}})\citenamefont
  {Xi}, \citenamefont {Xu},\ and\ \citenamefont {Diamond}}]{xi_phase_2014}%
  \BibitemOpen
  \bibfield  {author} {\bibinfo {author} {\bibfnamefont {P.}~\bibnamefont
  {Xi}}, \bibinfo {author} {\bibfnamefont {X.}~\bibnamefont {Xu}}, \ and\
  \bibinfo {author} {\bibfnamefont {P.}~\bibnamefont {Diamond}},\ }\href
  {\doibase 10.1103/PhysRevLett.112.085001} {\bibfield  {journal} {\bibinfo
  {journal} {Physical Review Letters}\ }\textbf {\bibinfo {volume} {112}},\
  \bibinfo {pages} {085001} (\bibinfo {year} {2014}{\natexlab{a}})}\BibitemShut
  {NoStop}%
\bibitem [{\citenamefont {Xi}\ \emph {et~al.}(2014{\natexlab{b}})\citenamefont
  {Xi}, \citenamefont {Xu},\ and\ \citenamefont {Diamond}}]{xi_impact_2014}%
  \BibitemOpen
  \bibfield  {author} {\bibinfo {author} {\bibfnamefont {P.~W.}\ \bibnamefont
  {Xi}}, \bibinfo {author} {\bibfnamefont {X.~Q.}\ \bibnamefont {Xu}}, \ and\
  \bibinfo {author} {\bibfnamefont {P.~H.}\ \bibnamefont {Diamond}},\ }\href
  {\doibase 10.1063/1.4875332} {\bibfield  {journal} {\bibinfo  {journal}
  {Physics of Plasmas}\ }\textbf {\bibinfo {volume} {21}},\ \bibinfo {pages}
  {056110} (\bibinfo {year} {2014}{\natexlab{b}})}\BibitemShut {NoStop}%
\bibitem [{\citenamefont {Gruzinov}\ and\ \citenamefont
  {Diamond}(1994)}]{gruzinov_self-consistent_1994}%
  \BibitemOpen
  \bibfield  {author} {\bibinfo {author} {\bibfnamefont {A.~V.}\ \bibnamefont
  {Gruzinov}}\ and\ \bibinfo {author} {\bibfnamefont {P.~H.}\ \bibnamefont
  {Diamond}},\ }\href {\doibase 10.1103/PhysRevLett.72.1651} {\bibfield
  {journal} {\bibinfo  {journal} {Physical Review Letters}\ }\textbf {\bibinfo
  {volume} {72}},\ \bibinfo {pages} {1651} (\bibinfo {year}
  {1994})}\BibitemShut {NoStop}%
\bibitem [{\citenamefont {Gruzinov}\ and\ \citenamefont
  {Diamond}(1996)}]{gruzinov_nonlinear_1996}%
  \BibitemOpen
  \bibfield  {author} {\bibinfo {author} {\bibfnamefont {A.~V.}\ \bibnamefont
  {Gruzinov}}\ and\ \bibinfo {author} {\bibfnamefont {P.~H.}\ \bibnamefont
  {Diamond}},\ }\href {\doibase 10.1063/1.871981} {\bibfield  {journal}
  {\bibinfo  {journal} {Physics of Plasmas}\ }\textbf {\bibinfo {volume} {3}},\
  \bibinfo {pages} {1853} (\bibinfo {year} {1996})}\BibitemShut {NoStop}%
\bibitem [{\citenamefont {Diamond}\ \emph {et~al.}(2005)\citenamefont
  {Diamond}, \citenamefont {Kim},\ and\ \citenamefont
  {Hughes}}]{diamond_self-consistent_2005}%
  \BibitemOpen
  \bibfield  {author} {\bibinfo {author} {\bibfnamefont {P.~H.}\ \bibnamefont
  {Diamond}}, \bibinfo {author} {\bibfnamefont {E.~J.}\ \bibnamefont {Kim}}, \
  and\ \bibinfo {author} {\bibfnamefont {D.~W.}\ \bibnamefont {Hughes}},\ }in\
  \href
  {https://books.google.com/books?id=PLNwoJ6qFoEC&lpg=PP1&pg=PA145#v=onepage&q&f=false}
  {\emph {\bibinfo {booktitle} {Fluid {Dynamics} and {Dynamos} in
  {Astrophysics} and {Geophysics}}}}\ (\bibinfo  {publisher} {CRC Press},\
  \bibinfo {year} {2005})\ p.\ \bibinfo {pages} {145}\BibitemShut {NoStop}%
\bibitem [{\citenamefont {Diamond}\ \emph {et~al.}(2010)\citenamefont
  {Diamond}, \citenamefont {Itoh},\ and\ \citenamefont
  {Itoh}}]{diamond_modern_2010}%
  \BibitemOpen
  \bibfield  {author} {\bibinfo {author} {\bibfnamefont {P.~H.}\ \bibnamefont
  {Diamond}}, \bibinfo {author} {\bibfnamefont {S.-I.}\ \bibnamefont {Itoh}}, \
  and\ \bibinfo {author} {\bibfnamefont {K.}~\bibnamefont {Itoh}},\ }\href
  {http://www.langtoninfo.com/web_content/9780521869201_frontmatter.pdf} {\emph
  {\bibinfo {title} {Modern {Plasma} {Physics}, {Physical} {Kinetics} of
  {Turbulence} {Plasmas} {Vol}. 1}}}\ (\bibinfo  {publisher} {Cambridge
  University Press},\ \bibinfo {year} {2010})\BibitemShut {NoStop}%
\bibitem [{\citenamefont {Zeldovich}(1957)}]{zeldovich_magnetic_1957}%
  \BibitemOpen
  \bibfield  {author} {\bibinfo {author} {\bibfnamefont {Y.~B.}\ \bibnamefont
  {Zeldovich}},\ }\href@noop {} {\bibfield  {journal} {\bibinfo  {journal}
  {Sov. Phys. JETP}\ }\textbf {\bibinfo {volume} {4}},\ \bibinfo {pages} {460}
  (\bibinfo {year} {1957})}\BibitemShut {NoStop}%
\bibitem [{\citenamefont {Pouquet}\ \emph {et~al.}(1976)\citenamefont
  {Pouquet}, \citenamefont {Frisch},\ and\ \citenamefont
  {Léorat}}]{pouquet_strong_1976}%
  \BibitemOpen
  \bibfield  {author} {\bibinfo {author} {\bibfnamefont {A.}~\bibnamefont
  {Pouquet}}, \bibinfo {author} {\bibfnamefont {U.}~\bibnamefont {Frisch}}, \
  and\ \bibinfo {author} {\bibfnamefont {J.}~\bibnamefont {Léorat}},\ }\href
  {http://journals.cambridge.org/abstract_S0022112076002140} {\bibfield
  {journal} {\bibinfo  {journal} {Journal of Fluid Mechanics}\ }\textbf
  {\bibinfo {volume} {77}},\ \bibinfo {pages} {321} (\bibinfo {year}
  {1976})}\BibitemShut {NoStop}%
\bibitem [{\citenamefont {Pouquet}(1978)}]{pouquet_two-dimensional_1978}%
  \BibitemOpen
  \bibfield  {author} {\bibinfo {author} {\bibfnamefont {A.}~\bibnamefont
  {Pouquet}},\ }\href {\doibase 10.1017/S0022112078001950} {\bibfield
  {journal} {\bibinfo  {journal} {Journal of Fluid Mechanics}\ }\textbf
  {\bibinfo {volume} {88}},\ \bibinfo {pages} {1} (\bibinfo {year}
  {1978})}\BibitemShut {NoStop}%
\bibitem [{\citenamefont {Ruiz}\ and\ \citenamefont
  {Nelson}(1981)}]{ruiz_turbulence_1981}%
  \BibitemOpen
  \bibfield  {author} {\bibinfo {author} {\bibfnamefont {R.}~\bibnamefont
  {Ruiz}}\ and\ \bibinfo {author} {\bibfnamefont {D.~R.}\ \bibnamefont
  {Nelson}},\ }\href
  {http://journals.aps.org/pra/abstract/10.1103/PhysRevA.23.3224} {\bibfield
  {journal} {\bibinfo  {journal} {Physical Review A}\ }\textbf {\bibinfo
  {volume} {23}},\ \bibinfo {pages} {3224} (\bibinfo {year}
  {1981})}\BibitemShut {NoStop}%
\bibitem [{\citenamefont {Fan}\ \emph {et~al.}(2016)\citenamefont {Fan},
  \citenamefont {Diamond}, \citenamefont {Chacón},\ and\ \citenamefont
  {Li}}]{fan_cascades_2016}%
  \BibitemOpen
  \bibfield  {author} {\bibinfo {author} {\bibfnamefont {X.}~\bibnamefont
  {Fan}}, \bibinfo {author} {\bibfnamefont {P.~H.}\ \bibnamefont {Diamond}},
  \bibinfo {author} {\bibfnamefont {L.}~\bibnamefont {Chacón}}, \ and\
  \bibinfo {author} {\bibfnamefont {H.}~\bibnamefont {Li}},\ }\href {\doibase
  10.1103/PhysRevFluids.1.054403} {\bibfield  {journal} {\bibinfo  {journal}
  {Physical Review Fluids}\ }\textbf {\bibinfo {volume} {1}},\ \bibinfo {pages}
  {054403} (\bibinfo {year} {2016})}\BibitemShut {NoStop}%
\bibitem [{\citenamefont {Fan}\ \emph {et~al.}(2017)\citenamefont {Fan},
  \citenamefont {Diamond},\ and\ \citenamefont {Chacón}}]{fan_formation_2017}%
  \BibitemOpen
  \bibfield  {author} {\bibinfo {author} {\bibfnamefont {X.}~\bibnamefont
  {Fan}}, \bibinfo {author} {\bibfnamefont {P.~H.}\ \bibnamefont {Diamond}}, \
  and\ \bibinfo {author} {\bibfnamefont {L.}~\bibnamefont {Chacón}},\ }\href
  {\doibase 10.1103/PhysRevE.96.041101} {\bibfield  {journal} {\bibinfo
  {journal} {Physical Review E}\ }\textbf {\bibinfo {volume} {96}},\ \bibinfo
  {pages} {041101} (\bibinfo {year} {2017})}\BibitemShut {NoStop}%
\bibitem [{\citenamefont {Fan}\ \emph {et~al.}(2018)\citenamefont {Fan},
  \citenamefont {Diamond},\ and\ \citenamefont {Chacón}}]{fan_chns:_2018}%
  \BibitemOpen
  \bibfield  {author} {\bibinfo {author} {\bibfnamefont {X.}~\bibnamefont
  {Fan}}, \bibinfo {author} {\bibfnamefont {P.~H.}\ \bibnamefont {Diamond}}, \
  and\ \bibinfo {author} {\bibfnamefont {L.}~\bibnamefont {Chacón}},\ }\href
  {\doibase 10.1063/1.5016075} {\bibfield  {journal} {\bibinfo  {journal}
  {Physics of Plasmas}\ }\textbf {\bibinfo {volume} {25}},\ \bibinfo {pages}
  {055702} (\bibinfo {year} {2018})}\BibitemShut {NoStop}%
\bibitem [{\citenamefont {Pandit}\ \emph {et~al.}(2017)\citenamefont {Pandit},
  \citenamefont {Banerjee}, \citenamefont {Bhatnagar}, \citenamefont {Brachet},
  \citenamefont {Gupta}, \citenamefont {Mitra}, \citenamefont {Pal},
  \citenamefont {Perlekar}, \citenamefont {Ray}, \citenamefont {Shukla},\ and\
  \citenamefont {Vincenzi}}]{pandit_overview_2017}%
  \BibitemOpen
  \bibfield  {author} {\bibinfo {author} {\bibfnamefont {R.}~\bibnamefont
  {Pandit}}, \bibinfo {author} {\bibfnamefont {D.}~\bibnamefont {Banerjee}},
  \bibinfo {author} {\bibfnamefont {A.}~\bibnamefont {Bhatnagar}}, \bibinfo
  {author} {\bibfnamefont {M.}~\bibnamefont {Brachet}}, \bibinfo {author}
  {\bibfnamefont {A.}~\bibnamefont {Gupta}}, \bibinfo {author} {\bibfnamefont
  {D.}~\bibnamefont {Mitra}}, \bibinfo {author} {\bibfnamefont
  {N.}~\bibnamefont {Pal}}, \bibinfo {author} {\bibfnamefont {P.}~\bibnamefont
  {Perlekar}}, \bibinfo {author} {\bibfnamefont {S.~S.}\ \bibnamefont {Ray}},
  \bibinfo {author} {\bibfnamefont {V.}~\bibnamefont {Shukla}}, \ and\ \bibinfo
  {author} {\bibfnamefont {D.}~\bibnamefont {Vincenzi}},\ }\href {\doibase
  10.1063/1.4986802} {\bibfield  {journal} {\bibinfo  {journal} {Physics of
  Fluids}\ }\textbf {\bibinfo {volume} {29}},\ \bibinfo {pages} {111112}
  (\bibinfo {year} {2017})}\BibitemShut {NoStop}%
\bibitem [{sup()}]{suppl_2019}%
  \BibitemOpen
  \href@noop {} {\bibinfo  {journal} {See Supplemental Material at [URL will be
  inserted by publisher] for details about the unimodal initial condition used
  in the main text, and the relationship between the CHNS system and MHD}\
  }\BibitemShut {NoStop}%
\bibitem [{\citenamefont {Chacon}\ \emph {et~al.}(2002)\citenamefont {Chacon},
  \citenamefont {Knoll},\ and\ \citenamefont {Finn}}]{chacon_implicit_2002}%
  \BibitemOpen
\bibfield  {journal} {  }\bibfield  {author} {\bibinfo {author} {\bibfnamefont
  {L.}~\bibnamefont {Chacon}}, \bibinfo {author} {\bibfnamefont
  {D.}~\bibnamefont {Knoll}}, \ and\ \bibinfo {author} {\bibfnamefont
  {J.}~\bibnamefont {Finn}},\ }\href {\doibase 10.1006/jcph.2002.7015}
  {\bibfield  {journal} {\bibinfo  {journal} {Journal of Computational
  Physics}\ }\textbf {\bibinfo {volume} {178}},\ \bibinfo {pages} {15}
  (\bibinfo {year} {2002})}\BibitemShut {NoStop}%
\bibitem [{\citenamefont {Chacon}\ and\ \citenamefont
  {Knoll}(2003)}]{chacon_2d_2003}%
  \BibitemOpen
  \bibfield  {author} {\bibinfo {author} {\bibfnamefont {L.}~\bibnamefont
  {Chacon}}\ and\ \bibinfo {author} {\bibfnamefont {D.}~\bibnamefont {Knoll}},\
  }\href {\doibase 10.1016/S0021-9991(03)00193-1} {\bibfield  {journal}
  {\bibinfo  {journal} {Journal of Computational Physics}\ }\textbf {\bibinfo
  {volume} {188}},\ \bibinfo {pages} {573} (\bibinfo {year}
  {2003})}\BibitemShut {NoStop}%
\bibitem [{\citenamefont {Ashourvan}\ and\ \citenamefont
  {Diamond}(2016)}]{ashourvan_how_2016}%
  \BibitemOpen
  \bibfield  {author} {\bibinfo {author} {\bibfnamefont {A.}~\bibnamefont
  {Ashourvan}}\ and\ \bibinfo {author} {\bibfnamefont {P.~H.}\ \bibnamefont
  {Diamond}},\ }\href {\doibase 10.1103/PhysRevE.94.051202} {\bibfield
  {journal} {\bibinfo  {journal} {Physical Review E}\ }\textbf {\bibinfo
  {volume} {94}},\ \bibinfo {pages} {051202} (\bibinfo {year}
  {2016})}\BibitemShut {NoStop}%
\bibitem [{\citenamefont {Ashourvan}\ and\ \citenamefont
  {Diamond}(2017)}]{ashourvan_emergence_2017}%
  \BibitemOpen
  \bibfield  {author} {\bibinfo {author} {\bibfnamefont {A.}~\bibnamefont
  {Ashourvan}}\ and\ \bibinfo {author} {\bibfnamefont {P.~H.}\ \bibnamefont
  {Diamond}},\ }\href {\doibase 10.1063/1.4973660} {\bibfield  {journal}
  {\bibinfo  {journal} {Physics of Plasmas}\ }\textbf {\bibinfo {volume}
  {24}},\ \bibinfo {pages} {012305} (\bibinfo {year} {2017})}\BibitemShut
  {NoStop}%
\end{thebibliography}%

\end{document}